# Computational evidence on repurposing the anti-influenza drugs baloxavir acid and baloxavir marboxil against COVID-19

Suban K Sahoo[*] and Seshu Vardhan

*Department of Applied Chemistry, S.V. National Institute of Technology (SVNIT), Surat-395007, India. (E-mail: sks@chem.svnit.ac.in, suban_sahoo@rediffmail.com)*

**Abstract:**

The main reasons for the ongoing COVID-19 (coronavirus disease 2019) pandemic are the unavailability of recommended efficacious drugs or vaccines along with the human to human transmission nature of SARS-CoV-2 virus. So, there is urgent need to search appropriate therapeutic approach by repurposing approved drugs. In this communication, molecular docking analyses of two influenza antiviral drugs baloxavir acid (BXA) and baloxavir marboxil (BXM) were performed with the three therapeutic target proteins of severe acute respiratory syndrome coronavirus 2 (SARS-CoV-2), i.e., main protease (Mpro), papain-like protease (PLpro) and RNA-dependent RNA polymerase (RdRp). The molecular docking results of both the drugs BXA and BXM were analysed and compared. The investigational drug BXA binds at the active site of Mpro and RdRp, whereas the approved drug BXM binds only at the active site of RdRp. Also, comparison of dock score revealed that BXA is binding more effectively at the active site of RdRp than BXM. The computational molecular docking revealed that the drug BXA may be more effective against COVID-19 as compared to BXM.

# 1. Introduction

The ongoing COVID-19 (coronavirus disease 2019) pandemic is caused by the virus strain severe acute respiratory syndrome coronavirus 2 (SARS-CoV-2) [1-3]. The virus was detected first at Wuhan, China in December, 2019, and as of 10$^{th}$ July 2020, globally total 12102328 confirmed COVID-19 cases were reported with the death of 551046. Till today, there is no recommended drugs or vaccines available to fight against the COVID-19. Therefore, to deal with this human to human transmissible SARS-CoV-2 virus, there is an emergent need to search potential drugs and/or phytochemicals that can be repurposed against the COVID-19 infection [4]. The Food and Drug Administration (FDA) approved drugs like remdesivir, favipiravir and tocilizumab etc. are under clinical trials to deal with the COVID-19 [5-7]. Most of the repurposing drugs are under initial phases of clinical trials and require further studies on their effectiveness and safety in the treatment of COVID-19. Therefore, the research work to search new potential drugs are also ongoing globally that can be repurposed against COVID-19.

For the drug repurposing research, the computer-based molecular docking and simulations studies were carried out to find the drug that bind effectively at the therapeutic target proteins of SARS-CoV-2 virus. SARS-CoV-2, the positive-sense single-stranded ribonucleic acid (ssRNA) virus is the new and seventh member of human coronaviruses (HCoVs) that has *beta*-coronavirus genus like severe acute respiratory syndrome coronavirus (SARS-CoV, first detected in 2003) and middle east respiratory syndrome coronavirus (MERS-CoV, first detected in 2012), with 26 to 32 kilobases and spheroid-shaped of 50-200 nm in diameter [8]. The virus has four important structural proteins, *i.e.,* spike (S), membrane (M), envelope (E) and nucleocapsid (N) proteins. The proteins S, M and E create the envelope of the SARS-CoV-2, where the club-shaped spike protein binds with the human host cells angiotensin converting enzyme 2 (ACE2) and initiates the viral entry [9]. Once the virus entered into host cells, the functional proteins like main protease (Mpro or 3CLpro), papain-like protease (PLpro) and RNA-dependent RNA polymerase (RdRp) play the vital role in viral replication and transcription [10]. Therefore, the proteins like ACE2, Mpro, PLpro, RdRp and spike glycoprotein are studied as therapeutic targets for the structure-based molecular docking and simulations to search potential inhibitors from the databases of approved drugs and phytochemicals.

Similar to COVID-19 symptoms, the influenza (flu) caused by influenza viruses is an acute respiratory infectious disease. The transmission of the flu viruses also occurs by contact, droplets and fomites. The heterotrimeric RNA polymerase of influenza virus composed of three protein subunits, polymerase acidic protein (PA), polymerase basic protein 1 (PB1), and polymerase basic protein 2 (PB2), that play a vital role in the viral replication. Based on a "cap-snatching" mechanism, the PB2 subunit interacts with the host nascent capped transcripts and subsequently cleaved by the cap-dependent endonuclease (CEN) present in PA subunit. The RNA primer generated from the "cap-snatching" mechanism is utilized by RNA-dependent RNA polymerase (RdRp) of PB1 for viral transcription [11]. The influenza antiviral drugs baloxavir marboxil (BXM) was approved in 2018 by FDA to treat influenza (**Fig. 1**), which can selectively inhibit the RNA replication activity of CEN [12,13]. The recent *in vitro* study of BXM revealed that the active form baloxavir acid (BXA) inhibits the viral RNA transcription by inhibiting the enzymatic activity of CEN selectively that ceases the viral replication in infected cells without any cytotoxicity [14]. More recent study on the antiviral drugs BXM and oseltamivir on influenza revealed that the BXM prevents transmission by blocking the virus replication more effectively and completely than the oseltamivir [15]. Despite the potential inhibiting ability of BXM and BXA on influenza treatment, there are only a few clinical trials on COVID-19 [16,17].

Considering the need of new potential drugs to repurpose against COVID-19 pandemic, this research was carried out to investigate the effective binding of the drugs BMX and BXA with the three functional proteins of SARS-CoV-2, i.e., Mpro, PLpro and RdRp. The molecular docking of the drugs were performed with the therapeutic target proteins and their affinity of binding at the active site was compared.

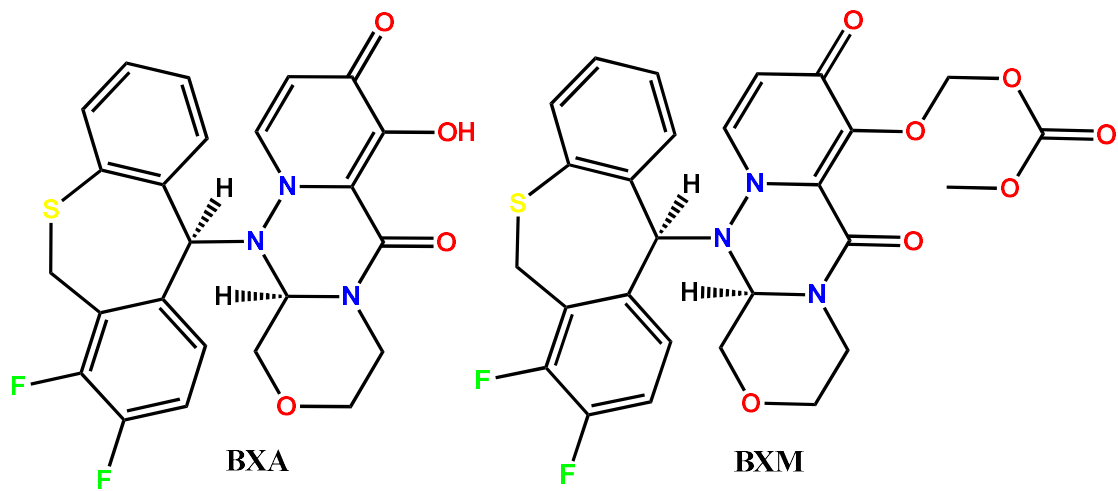

**Fig. 1.** Chemical structure of the anti-influenza drugs baloxavir acid (BXA) and baloxavir marboxil (BXM).

## 2. Experimental

The molecular docking analyses were done in AutoDock Vina [18] to screen the most effective drugs between the BXA and BXM. The 3D structure of the drugs were downloaded from the PubChem (https://pubchem.ncbi.nlm.nih.gov/). The crystal structures of the proteins Mpro (PDB ID: 6LU7), PLpro (PDB ID: 4MM3) and RdRp (PDB ID: 6M71) were retrieved from the PDB database (www.rcsb.org). These proteins were modelled by using the protein template of pdb structure from the online server SWISS-MODEL. The proteins and drugs were prepared for molecular docking by utilizing the AutoDock tools. The grid box dimensions (55.59 Å × 72.96 Å × 75.01 Å) centered at (-24.839, 13.571, 58.159) for the Mpro, the grid box dimensions (55.59 Å × 72.96 Å × 75.01 Å) centered at (-26.097, 19.781, 27.145) for the PLpro, and the grid box dimensions (77.17 Å × 88.61 Å × 101.09 Å) centered at (120.50, 119.46, 128.42) for the RdRp were generated to cover the whole protein structure. Total ten poses with the best docking score were saved and the output files were analysed using the visualisation software BioVia Discovery Studio.

## 3. Results and discussion

With the advancement of computational facilities and the accuracy of the prediction, the computer-based molecular docking gained a huge popularity in the field of drugs design and discovery. Molecular docking can determine the binding affinity and binding poses at the active site of the target proteins. The binding affinity referred as the dock score determine the stability of the protein-ligand complex formed, whereas the binding poses at the active site will provide evidence on the efficacy of the ligand to inhibit the enzymatic activity [19]. Therefore, to provide computational evidence on the comparative potency of the two influenza antiviral drugs BXA and BXM (**Fig. 1**), the molecular docking experiments were performed in Autodock Vina with the therapeutic target proteins of SARS-CoV-2, *i.e.,* Mpro, PLpro and RdRp.

The blind molecular docking was performed where the grid box was selected to cover the whole protein structure. The ten best dock score poses of the drugs BXA and BXM with the proteins Mpro, PLpro and RdRp were saved, and the protein-ligand interaction study was performed. The drugs BXA and BXM bind with the protein PLpro with the dock score ranging from -8.9 to -7.2 kcal/mol and -7.0 to -6.5 kcal/mol respectively, but the *in silico* protein-ligand interaction study revealed that no binding poses found at the active site consists of the three catalytic triad (CYS112, HIS273 and ASP287) of the PLpro [20]. Therefore, the enzymatic activity of PLpro is not inhibited by the BXA and BXM. With the Mpro, analyses of the ten best binding poses of BXA and BXM revealed that the drug BMX with dock score ranging from -8.3 to -6.5 kcal/mol failed to bind at the active site present at the catalytic dyad CYS145 and HIS41 [21]. In contrast, the drug BXA binds at the active site of Mpro with the estimated dock score of -7.4 kcal/mol and therefore expected to interfere with the activity of Mpro. The posing ligand BXA binds exclusively at the target chain of Mpro where the active site is located (**Fig. 2**). BXA formed Van der Waals (VDW) and pi-alkyl interactions with the catalytic dyad of Mpro HIS41 and CYS145, respectively. The effective binding between BXA and Mpro can also be examined from the multiple non-covalent bindings with other residues at the active site (**Fig. 2a-b**). BXA formed VDW bonds with the residues ASP187, HIS164, GLY143, SER144, HIS163 and HIS172. The residues ARG188 and LEU141 are involved in the carbon hydrogen bond. The fluorine atoms of BXA are participated in halogen bonds with the residues PHE140, ASN142 and GLU166. The alkyl and pi-alkyl bonds are observed with the residues MET49 and MET165. In addition, the sulphur atom of BXA also participates in the non-covalent interaction with the GLN189. Further, the 3D binding pose of BXA at the active site of Mpro supporting the high stability of the BXA-Mpro complex (**Fig. 2c**). It is also important to mention here that the binding energy of BXA is comparably higher than the drugs like oseltamivir (-4.7 kcal/mol), ritonavir (-7.3 kcal/mol), remdesivir (-6.5 kcal/mol), ribavirin (-5.6 kcal/mol), favipiravir (-5.4 kcal/mol), chloroquine (-5.1 kcal/mol) and hydroxychloroquine (-5.3 kcal/mol) [22].

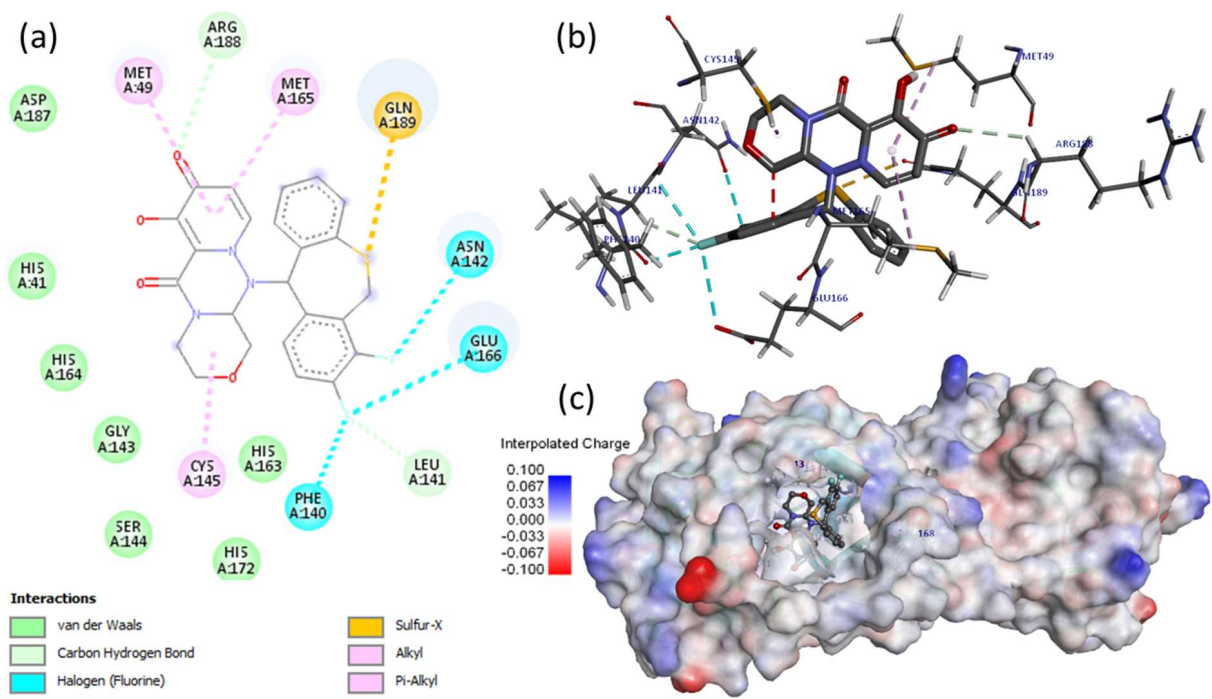

**Fig. 2.** (a) 2D and (b) 3D poses showing the interactions between BXA and Mpro, and (3) 3D docked pose showing the inclusion of BXA within the cavity of Mpro.

The drugs BXA and BXM bind with the SARS-CoV-2 RdRp protein with the dock score ranging from -9.3 to -8.1 kcal/mol and -9.0 to -7.4 kcal/mol, respectively. The SARS-CoV-2 RdRp plays major role in the viral transcription and replication, and therefore RdRp is the most promising therapeutic target protein for searching repurpose drugs. The modelled SARS-CoV-2 RdRp protein structure consists of nsp12 that bound to nsp7 and nsp8 domains. In nsp12 [23], the RNA polymerase domain consists of a finger subdomain residues from 366 to 581 and 621 to 679, a palm subdomain residues from 582 to 620 and 680 to 815, and a thumb subdomain residues from 816 to 920. These polymerase residues formed the catalytic pocket of SARS-CoV-2 RdRp. The protein-ligand interaction study revealed that both the drugs BXA and BXM are binding at the active site of SARS-CoV-2 RdRp protein with the dock score of -8.1 and -7.6 kcal/mol, respectively. The docking results supporting the ability of the influenza antiviral drugs BXA and BXM to inhibit the activity of SARS-CoV-2 RdRp protein. Based on the dock score, the effectiveness of the investigational drug BXA is comparably better than the approved drug BXM to inhibit the activity of SARS-CoV-2 RdRp. In addition, the dock score of BXA and BXM at the active site of RdRp is comparably better that the antiviral drugs like remdesivir and ribavirin [24].

The docking pose of BXA with the RdRp protein is shown in **Fig. 3**. BXA is forming hydrogen bond with the residue ARG569. VDW interactions are observed with the residues ILE589, GLY683, LYS500, TYR689, ASP684 and LYS577. Residues ALA688 and LEU576 are forming pi-alkyl interactions with BXA. BXA also formed the halogen and carbon hydrogen bonds with the residues SER682 and ALA685, respectively. The 3D dock pose of BXA-RdRp complex indicating that the drug BXA is well buried inside the catalytic pocket and the complex structure is stabilized by multiple non-covalent interactions (**Fig. 3b-c**). Similar to BXA, the BXM-RdRp complex structure is also stabilized by various non-covalent bonds (**Fig. 4**). The residues ARG569 and GLN573 formed conventional hydrogen bonds with BXM. The VDW interactions were observed with the residues ILE589, ASN496, ASN497, ILE494 and ALA581. The residue TRY689 formed pi-pi interaction, whereas the pi-alkyl interactions were observed with the residues ALA688, LEU576, ALA685 and ALA580. BXM also formed a carbon hydrogen bond with the residue LYS577 at the active site.

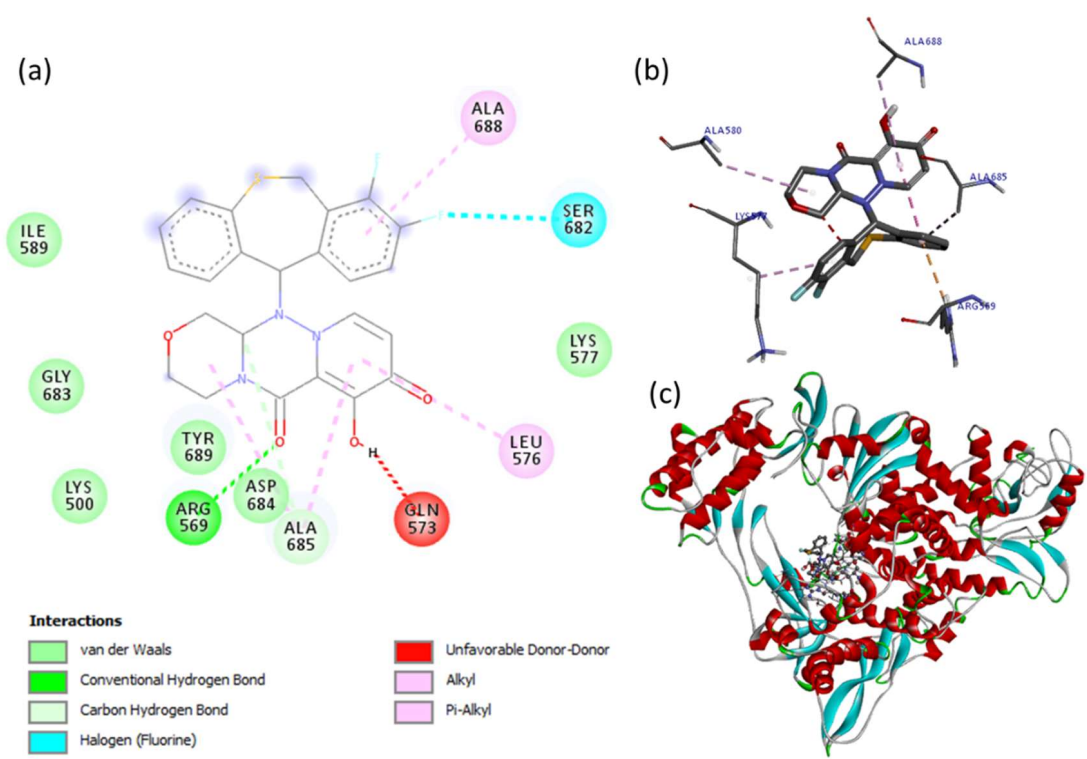

**Fig. 3. (**a) 2D and (b) 3D poses showing the interactions between BXA and RdRp, and (3) 3D docked pose showing the inclusion of BXA within the cavity of RdRp.

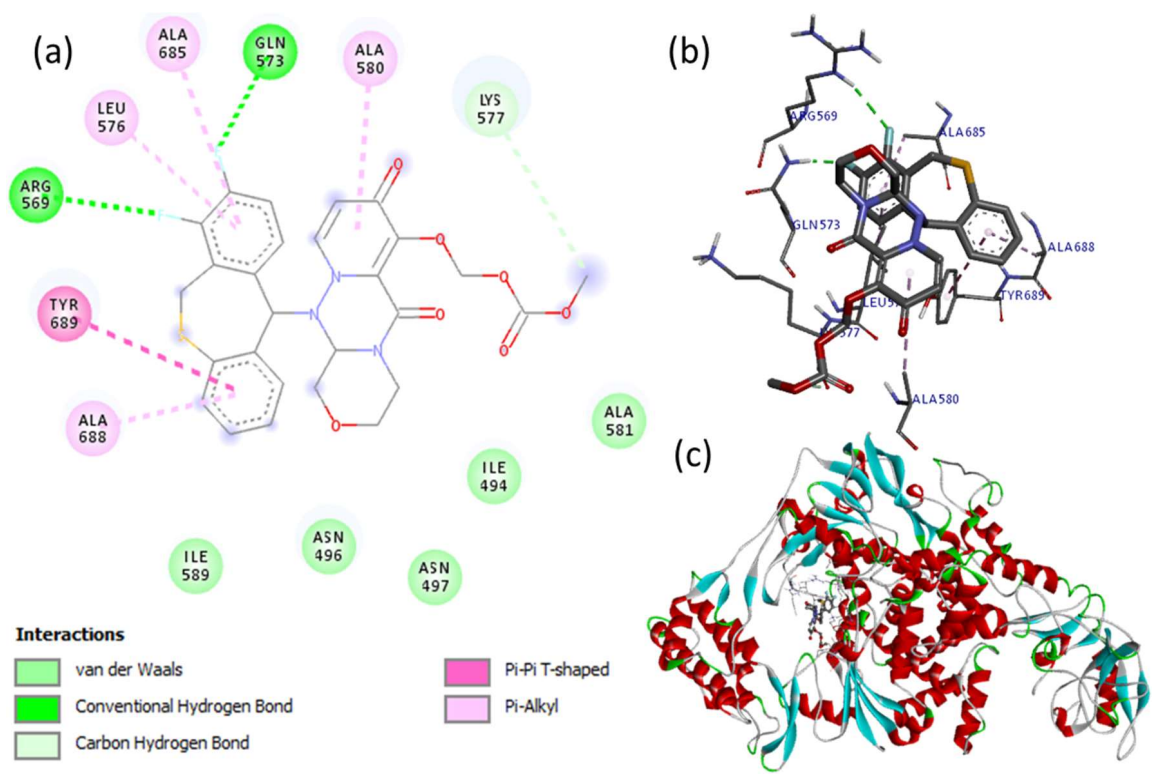

**Fig. 4. (**a) 2D and (b) 3D poses showing the interactions between BXM and RdRp, and (3) 3D docked structure showing the inclusion of BXM within the cavity of RdRp.

Finally, we additionally performed the site-specific docking of the drugs by generating the grid box dimensions (40 Å × 40 Å × 54 Å) centered at (-14.170, 39.730, -39.510) at active site residue VAL557 and surrounding amino acid residues for the RdRp. The top five poses of BXA showed the dock score of -8.2, -8.0, -7.8, -7.6 and -7.5 kcal/mol, whereas the drug BXM showed -7.7, -7.7, -7.6. -7.6 and -7.4 kcal/mol. The dock score and the protein-ligand interaction study of the top five poses indicating the effective binding of the drugs at the active site of SARS-CoV-2 RdRp. Also, the comparison of dock score clearly delineated that the drug BXA is showing better binding affinity at the active site as compared to BXM.

**4. Conclusions**

In summary, we have performed computer-based molecular docking of the new influenza antiviral drugs BXA and BXM with the three important therapeutic target proteins of SARS-CoV-2. The investigational drug BXA binds at the active site of main protease and RdRp, whereas the approved drug BXM binds only at the active site of RdRp. Also, comparison of dock score revealed that BXA is binding more effectively at the active site of RdRp than BXM. Therefore, we conclude that BXA may be more effective against COVID-19 as compared to BXM. Finally, we believe the current outcomes will be useful in formulating new therapeutic approach to fight against COVID-19.

**Declaration of competing interest**

The authors declare that they have no known competing financial interests or personal relationships that could have appeared to influence the work reported in this paper.

**Acknowledgments**

Authors are thankful to Direct, SVNIT for providing necessary research facilities and infrastructure.

**Data availability**

The docking structures can be obtained from the corresponding author upon request.

**References**

1. R. Hussin, et al., The epidemiology and pathogenesis of coronavirus disease (COVID-19) Outbreak. J. Autoimmun. 109 (2020) 102433.

2. N. Zhu, et al., A Novel Coronavirus from Patients with Pneumonia in China, 2019. N. Engl. J. Med. 382 (2020) 727-733.

3. C. Liu, et al., Research and Development on Therapeutic Agents and Vaccines for COVID-19 and Related Human Coronavirus Diseases. ACS Cent. Sci. 6 (2020) 315-331.

4. S. Vardhan, et al., *In silico* ADMET and molecular docking study on searching potential inhibitors from limonoids and triterpenoids for COVID-19. Comput. Bio. Med. 124 (2020) 103936.

5. K.-T. Choy, et al., Remdesivir, lopinavir, emetine, and homoharringtonine inhibit SARS-CoV-2 replication in vitro. Antivir. Res. 178 (2020) 104786.

6. L. Dong, et al., Discovering drugs to treat coronavirus disease 2019 (COVID-19). Drug Discov. Ther. 14 (2020) 58-60.

7. D. McKee, et al., Candidate drugs against SARS-CoV-2 and COVID-19. Pharmacol. Res. 157 (2020) 104859.

8. J. Zheng, SARS-CoV-2: an Emerging Coronavirus that Causes a Global Threat. Int. J. Biol. Sci. 16 (2020) 1678-1685.

9. D. Kim, et al., The Architecture of SARS-CoV-2 Transcriptome. Cell 181 (2020) 914-921.

10. A.C. Walls, et al., Structure, Function, and Antigenicity of the SARS-CoV-2 Spike Glycoprotein. Cell 180 (2020) 281-292.

11. F.G. Hayden, et al., Baloxavir Marboxil for Uncomplicated Influenza in Adults and Adolescents. N. Engl. J. Med. 379 (2018) 913-923.

12. Y.-A. Heo, Baloxavir: First Global Approval. Drugs 78 (2018) 693-697.

13. R. Yoshino, et al., Molecular Dynamics Simulation reveals the mechanism by which the Influenza Cap-dependent Endonuclease acquires resistance against Baloxavir marboxil. Sci. Rep. 9 (2019) 17464.

14. T. Noshi, et al., In vitro characterization of baloxavir acid, a first-in-class cap-dependent endonuclease inhibitor of the influenza virus polymerase PA subunit. Antivir. Res. 160 (2018) 109-117.

15. Z. Du, et al., Modeling mitigation of influenza epidemics by baloxavir. Nat. Commun. 11 (2020) 2750.

16. Y. Lou, et al., Clinical Outcomes and Plasma Concentrations 1 of Baloxavir Marboxil and Favipiravir in COVID-19 Patients: An Exploratory Randomized, Controlled Trial. medRxiv (2020) 2020.04.29.20085761.

17. G. Li, et al., Therapeutic options for the 2019 novel coronavirus (2019-nCoV), Nat. Rev. 19 (2020) 149-150.

18. O. Trott, et al., AutoDock Vina: Improving the speed and accuracy of docking with a new scoring function, efficient optimization, and multithreading. J. Comput. Chem. (2009) 455-461.

19. M. Kandeel, et al., Virtual screening and repurposing of FDA approved drugs against COVID-19 main protease. Life Sci. 251 (2020) 117627.

20. Y. Baez-Santos, et al., Catalytic Function and Substrate Specificity of the Papain-Like Protease Domain of nsp3 from the Middle East Respiratory Syndrome Coronavirus. J. Virol. 88 (2014) 12511-12527.

21. Z. Jin, et al., Structure of Mpro from COVID-19 virus and discovery of its inhibitors. Nature 582 (2020) 289–293.

22. R.R. Narkhede, et al., The Molecular Docking Study of Potential Drug Candidates Showing Anti-COVID-19 Activity by Exploring of Therapeutic Targets of SARS-CoV-2. EJMO 4 (2020) 185-195.

23. Y. Gao, et al., Structure of the RNA-dependent RNA polymerase from COVID-19 virus. Science 368 (2020) 779-782.

24. R. Yu, et al., Computational screening of antagonists against the SARS-CoV-2 (COVID-19) coronavirus by molecular docking. Int. J. Antimicrob. Agents (2020) https://doi.org/10.1016/j.ijantimicag.2020.106012.